\documentclass{ctextemp_elsart}

\usepackage{graphicx}

\usepackage{amssymb}

\usepackage{amsmath}

\begin{document}

\begin{frontmatter}



\title{Note on two phase phenomena in financial markets}


\author[label1]{Shi-Mei Jiang}
\author[label1]{Shi-Min Cai}
\author[label1,label2,label3]{Tao Zhou*}\corauth[cor1]{Email address: zhutou@ustc.edu}
\author[label1]{Pei-Ling Zhou}

\address[label1]{Department of Electronic Science and
Technology, University of Science and Technology of China, Hefei
Anhui, 230026, PR China}
\address[label2]{Department of Modern Physics,
University of Science and Technology of China, Hefei Anhui, 230026,
PR China}
\address[label3]{Department of Physics, University of
Fribourg, Chemin du Muse 3, CH-1700 Fribourg, Switzerland}

\begin{abstract}
The two phase behavior in financial markets actually means the
bifurcation phenomenon, which represents the change of the
conditional probability from an unimodal to a bimodal distribution.
In this paper, the bifurcation phenomenon in Hang-Seng index is
carefully investigated. It is observed that the bifurcation
phenomenon in financial index is not universal, but specific under
certain conditions. The phenomenon just emerges when the power-law
exponent of absolute increment distribution is between 1 and 2 with
appropriate period. Simulations on a randomly generated time series
suggest the bifurcation phenomenon itself is subject to the
statistics of absolute increment, thus it may not be able to reflect
the essential financial behaviors. However, even under the same
distribution of absolute increment, the range where bifurcation
phenomenon occurs is far different from real market to artificial
data, which may reflect certain market information.
\end{abstract}
\end{frontmatter}


\hspace{0.5cm}Financial markets are typical complex systems. To
understand their dynamics requires interdisciplinary knowledge and
exploration, including the application of concepts and tools of
statistical physics. Since the early 70's, a number of physicists
have devoted their time to the study of economic and financial
phenomena
\cite{Anderson1988,Arthur1997,Kertesz1997,Mantegna1999,Johnson2003,Challet2005}.
They have developed a wide range of concepts and models, including
fractal and multifractal scaling, frustrated disordered systems,
phenomena far from equilibrium, and so on
\cite{Bouchaud2000,Mandelbrot1997,Sornette2000,Wang2001,Cai2006,SMCai2006}.

\hspace{0.5cm}Recently, using transactions and quotes data for 116
most-actively traded US stocks for the 2 yr period 1994-1995,
Plerou-Gopikrishnan-Stanley empirically discovered a two-phase
behavior in financial markets \cite{Plerou2003}. Introducing a
parameter $\Sigma$ describing the fluctuation during the time
interval $\Delta t$, the conditional probability distribution
$p(\Omega \arrowvert \Sigma)$ of the volume imbalance $\Omega$, is
found to be with a single peak for $\Sigma < \Sigma_c $ and double
peaks for $\Sigma > \Sigma_c $. At the critical value $\Sigma_c$,
the transition from a single peak to double peaks occurs. The change
of $p(\Omega \arrowvert \Sigma)$ from an unimodal to a bimodal
distribution (the bifurcation phenomenon) indicates that the market
moves between an `equilibrium' state and an `out-of-equilibrium'
state, and these two different states were interpreted as distinct
phases. Following this idea, Zheng \emph{et al.} \cite{Zheng2004}
investigated the bifurcation phenomenon in financial markets with
the time series of the German DAX from 1994 to 1997. It was observed
that the probability distribution of the return $Z(t)$ conditioned
on the fluctuation of the financial index $r(t)$ displays a
transition from a unimodal distribution for small $r$, to a bimodal
distribution for large $r$. However, some recent works on trading
volume indicate that the bifurcation phenomenon is an artifact of
the distribution of trade sizes $q$, which follows a power-law
distribution with exponent $\zeta_{q}<2$ in the L$\acute{e}$vy
stable domain \cite{potter,Plerou2005,Matia2005}.

\hspace{0.5cm}In this paper, the bifurcation phenomenon is
investigated with the minute-by-minute records of Hang-Seng index
from 1 July 1994 to 28 May 1997 (see Fig. 1 the index and its
absolute increment). The trading time for a trading day in the data
was not the same in the whole period. Although for all trading days,
the Hong Kong stock market opened from 10:00 A.M. to 12:30 P.M. for
the morning session, occupying a time interval of 150min and opened
from 2:30 P.M. in the afternoon, the closing times were not the
same. From 1 July 1994 to 30 August 1995, the market closed at 3:45
P.M. with the total trading time 225 min per day. From 1 September
1995 to 30 December 1996, the market closed at 3:55 P.M. with the
total trading time 235min per day. From 1 January 1997 to 28 May
1997, the market closed at 4:00 P.M. with the total trading time 240
min per day. The total number of data points is 165727. In order to
carefully investigate the character of the Hang-Seng index, the
total data is divided into six segments every half a year, as shown
in Fig. 1(b). The largest number of data points in one segment is
29889, while the smallest is 24442. The number of data points is
sufficiently large for a detailed statistical analysis.

\begin{figure}
\scalebox{0.8}[0.8]{\includegraphics{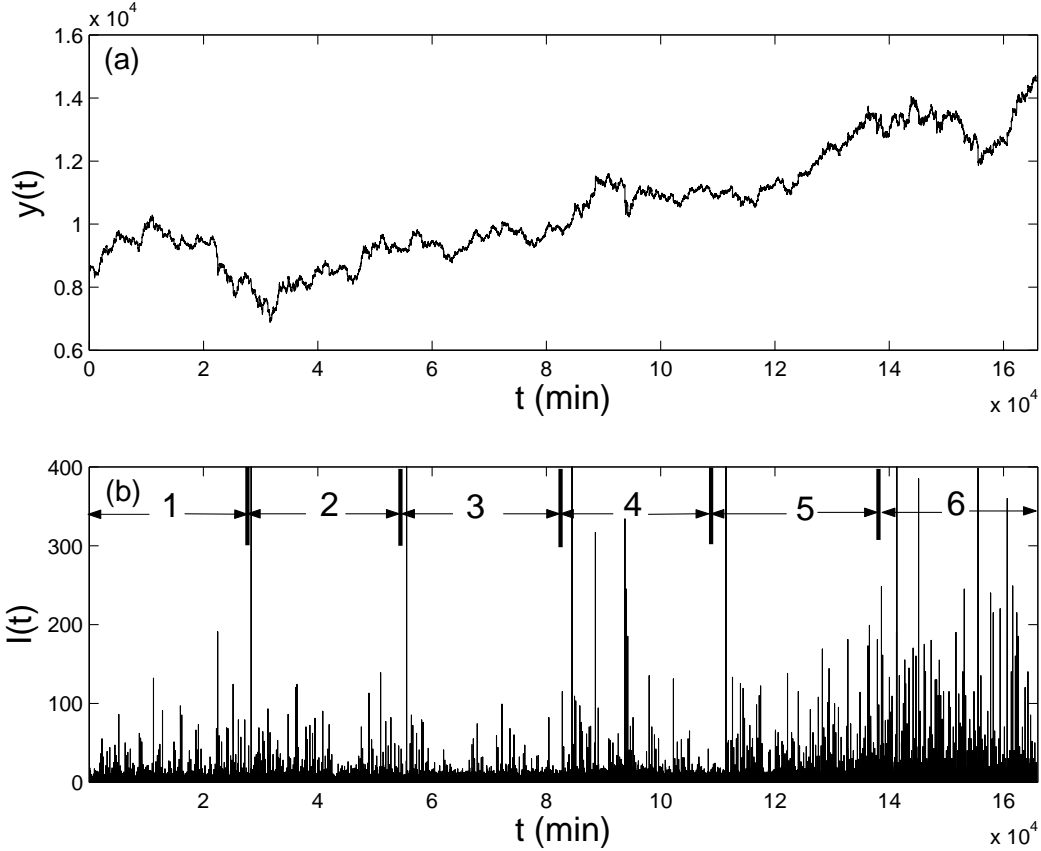}}
\caption{\label{fig:epsart}(a) The Hang-Seng index from 1 July
1994 to 28 May 1997 at the sampling intervals 1 minute. (b) The
corresponding absolute increment $I(t)$.}
\end{figure}

\hspace{0.5cm}Denote by $y(t)$ the time series of the Hang-Seng
index, the corresponding absolute increment reads
\begin{equation}
I(t)=\mid y(t+1)-y(t)\mid.
\end{equation}
The fluctuation $r_{\Delta t}(t)$ is simply the relative variation
from $t$ to $t+\Delta t$
\begin{equation}
r_{\Delta t}(t)=\langle\mid y(t+1)-y(t)-\langle
y(t+1)-y(t)\rangle_{\Delta t}\mid\rangle_{\Delta t},
\end{equation}
where $\langle\rangle_{\Delta t}$ denotes the average from $t$ to
$t+\Delta t$, and $y(t+1)$ means $y(t+1min)$. For a fixed $\Delta
t$, we calculate the conditional probability distribution $p_{\Delta
t}(Z,r)=p_{\Delta t}(Z|r)$ of the return $Z(t)=y(t+\Delta t)-y(t)$
with a specified $r$.

\begin{table}[table1]
\begin{center}
\caption{Exponents and the existence of bifurcation phenomenon.}

\begin{tabular}{ccc}\hline

segment ID & exponent & unimodal to bimodal\\
\hline 1 & 2.12 $\pm$ 0.03 & N\\
\hline 2 & 2.32 $\pm$ 0.04 & N\\
\hline 3 & 2.44 $\pm$ 0.02 & N\\
\hline 4 & 2.07 $\pm$ 0.03 & N\\
\hline 5 & 2.09 $\pm$ 0.03 & N\\
\hline 6 & 1.93 $\pm$ 0.03 & Y\\ \hline
\end{tabular}
\end{center}
\end{table}

\hspace{0.5cm}Fig. 2(a)-(f) show the empirical results of six
segments, respectively. It can be found that for all the six
segments, when the fluctuation $r$ is very small, the distribution
of return is single-peaked at about zero. However, for segment one
to five, different from the expected two phase phenomena
\cite{Plerou2003}, when the fluctuation gets bigger the distribution
of return is not a bimodal distribution. Instead, each of those five
has more than two maxima. Actually, clear transition from unimodal
to bimodal distribution can not be observed with the scale ranging
from 1 min up to about a day. In contrast, as shown in Fig. 2(f),
when the fluctuation $r<4.8$ the distribution of return $p_{\Delta
t}(Z,r)$ is single-peaked; when the fluctuation $r>4.8$, the bigger
the fluctuation is, the clearer the bimodal distribution becomes.
The transition from unimodal to bimodal distribution holds for the
scale ranging from 75min to 125min. Beyond the range this phenomenon
fades away.

\begin{figure}
\scalebox{1.2}[1.2]{\includegraphics{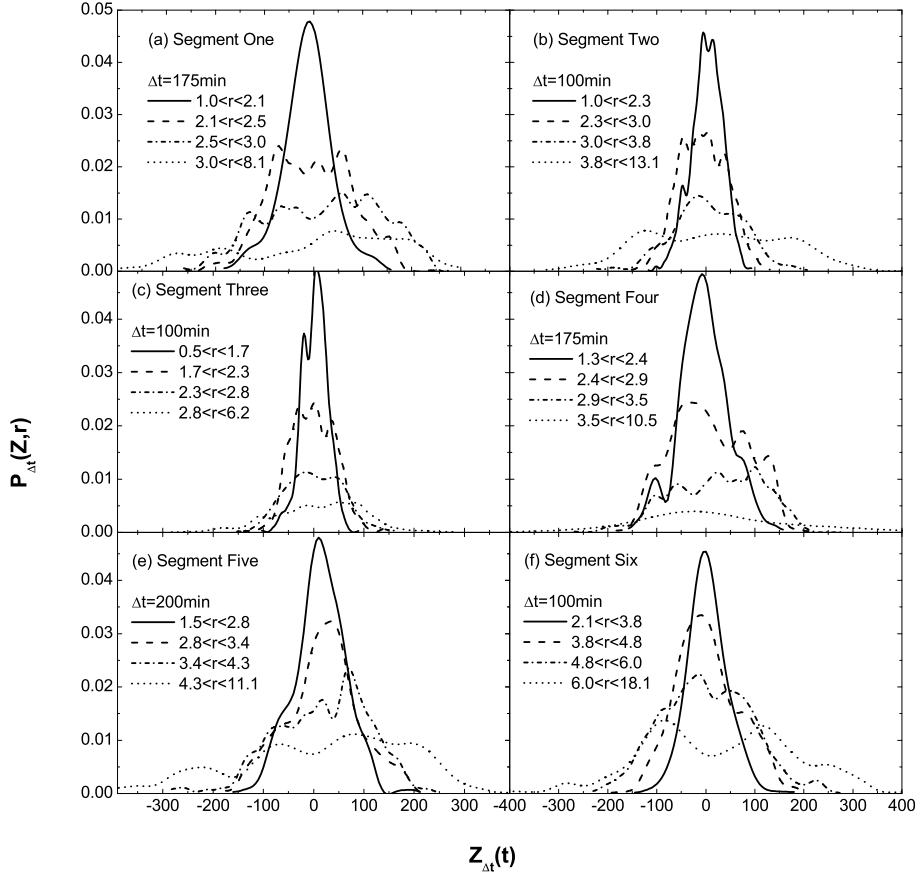}}
\caption{\label{fig:epsart} The distribution $p_{\Delta t}(Z,r)$
of six segments: (a) segment one from 1 July to 30 December 1994
with the scale 150 min, (b) segment two from 3 January to 30 June
1995 with the scale 100 min, (c) segment three from 3 July to 29
December 1995 with the scale 100 min, (d) segment four from 2
January to 28 June 1996 with the scale 175 min, (e) segment five
from 1 July to 31 December 1996 with the scale 200 min, (f)
segment six from 2 January to 28 June 1997 with the scale 100
min.}
\end{figure}

\hspace{0.5cm}The study on trading volume shows that the transiton
of $p(\Omega \arrowvert \Sigma)$ from an unimodal to a bimodal
distribution is an artifact of the distribution of trade sizes $q$,
which obeys a power-law distribution with exponent $ \zeta_{q}<2$ in
the L$\acute{e}$vy stable domain \cite{potter,Plerou2005,Matia2005}.
Similar to the consideration in Refs. \cite{Plerou2005,Matia2005},
we guess the existence of bifurcation phenomenon of stock index is
dependent on the statistics of absolute increment $I$. Two typical
cumulative distributions of $I$ are reported in Fig. 3, which both
follow a power-law form above a lower bound $I_{min}$. To
demonstrate the stability of the distributions, PDFs of return for
different time scales are analyzed. As an example, Fig. 4 shows the
re-scaled distributions for segment six. From Fig. 4 one can observe
that the distributions for different time scales well collapse onto
one master curve, which implies the stability of the distribution.
We use the method of the best-fit power-law model and
Kolmogorov-Smirnov (KS) statistic
\cite{kolmogorov,Sinai,Clauset2007} to estimate parameters in the
distribution, including both the lower bound $I_{min}$ and the
power-law exponent $\zeta_{I}$. In order to estimate carefully and
accurately, we also apply the usual method of least-squares (LS) on
the logarithm of the histogram. The exponents obtained by those two
methods are nearly the same (see Fig. 3), and we report the average
value (see Table 1). In Table 1, `N' means no bifurcation phenomenon
appears and `Y' means the phenomenon can be observed. Compared with
other five segments, the exponent of segment six is the smallest.
Accordingly, we guess the bifurcation phenomenon can be observed
only when $\zeta_{I}<2$.

\hspace{0.5cm}Furthermore, given a power-law distribution of $I$,
we generate artificial absolute increment time series $I(t)$ using
the method introduced in Ref. \cite{Clauset2007}. The sign of
increment (could be $+$ or $-$) is randomly assigned, that is to
say, the increment $i(t)$ is equal to $I(t)$ or $-I(t)$.
Accordingly, $z(t)$ and $r(t)$ are
\begin{equation}
z(t)=\sum^{t+\Delta(t)}_{\tau=t}i(\tau),
\end{equation}
and
\begin{equation}
r_{\Delta t}(t)=\langle\mid i(t)-\langle i(t)\rangle_{\Delta
t}\mid\rangle_{\Delta t}.
\end{equation}
The number of data points and the lower bound are set to be same as
the real ones. The bifurcation phenomenon is clearly observed in
Fig. 5(a) and 5(c), which holds for the scale ranging narrowly from
3 to 15 and from 3 to 18, respectively. Artificial data obeying
power-law distribution with exponents ranging from 0 to 3 are
carefully investigated, it is found that the obvious bifurcation
phenomenon only holds when the power-law exponent $\zeta_{I}$
satisfies $1<\zeta_{I}<2$. There is no bifurcation phenomenon with
$\zeta_{I}>2$ and $0<\zeta_{I}<1$ whatever the scale is (see, for
example, Fig. 5(b) and 5(d)).

\begin{figure}
\scalebox{0.8}[0.8]{\includegraphics{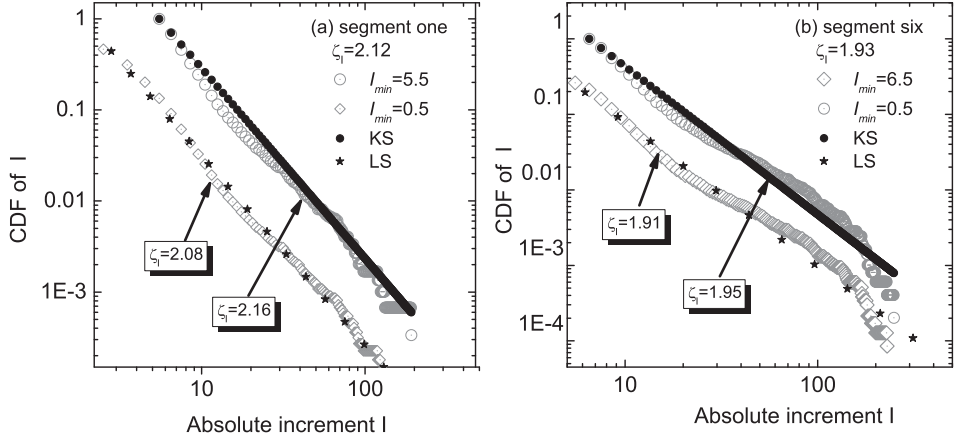}}
\caption{\label{fig:epsart} The cumulative probability of absolute
increment. (a) Segment one with average exponent $\zeta_{I}=2.12 \pm
0.03$. (b) Segment six with average exponent $\zeta_{I}=1.93 \pm
0.03 $. CDF stands for the cumulative distribution function.}
\end{figure}

\begin{figure}
\center \scalebox{0.8}[0.8]{\includegraphics{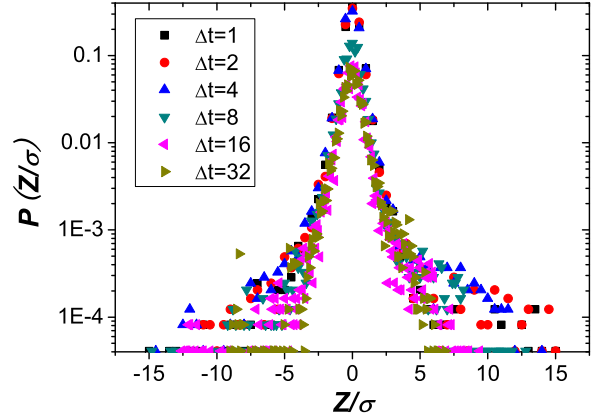}}
\caption{\label{fig:epsart} (Color online) Re-scaled plot of the
probability distributions. The abscissa is the re-scaled returns,
and the ordinate is the logarithm of re-scaled probability.}
\end{figure}

\begin{figure}
\scalebox{1.2}[1.2]{\includegraphics{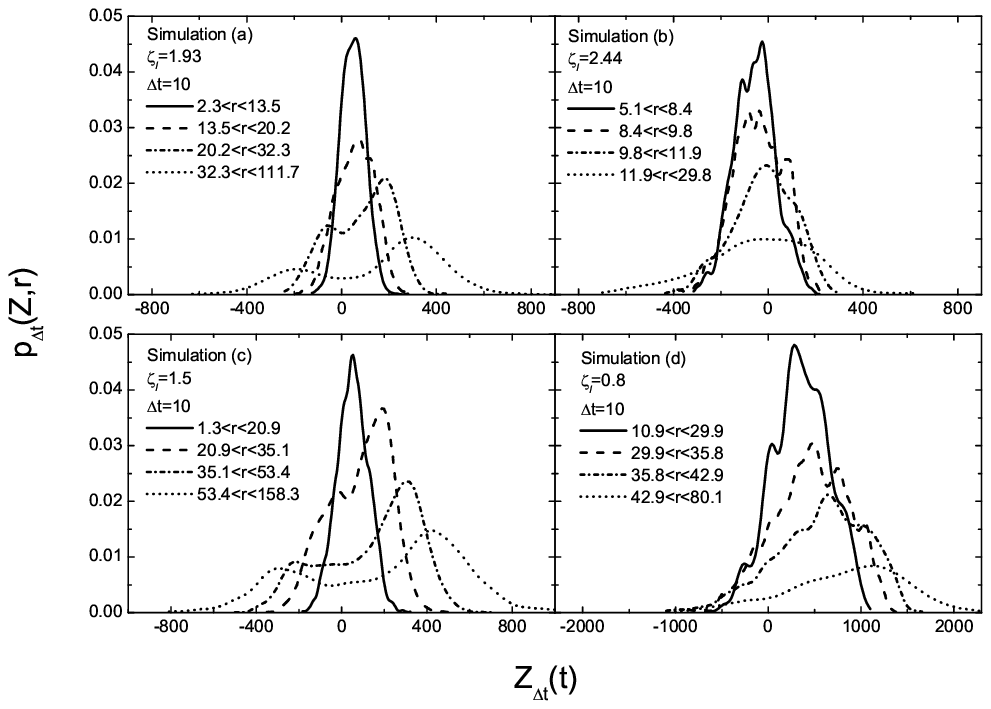}}
\caption{\label{fig:epsart} Numerical simulation of the bifurcation
phenomenon on artificial data. Power-law distributed $I$ with the
parameters: (a) $I_{min}=7$, $\zeta_{I}=1.93$, 24442 data points as
same as segment six; (b) $I_{min}=5$, $\zeta_{I}=2.44$, 28955 data
points as same as segment three; (c) $I_{min}=7$, $\zeta_{I}=1.50$,
25000 data points; (d) $I_{min}=7$, $\zeta_{I}=0.80$, 25000 data
points.}
\end{figure}

\hspace{0.5cm}Our findings suggest that the bifurcation phenomenon
in financial index is not universal, but specific under certain
conditions. The phenomenon just happens, within an appropriate
period of time scale, when the power-law exponent of absolute
increment distribution is between 1 and 2. The simulations on
randomly generated time series suggest the bifurcation phenomenon
itself is subject to the statistics of absolute increment, thus it
may not be able to reflect the essential financial behaviors (see
also the relative comments from Refs. \cite{potter,Krzesinski}).
However, one should note that, even under the same distribution of
absolute increment, the range where bifurcation phenomenon occurs is
far different from real market to artificial data: for actual index
the appropriate period is wide, while for the artificial data, it is
very narrow (compare Fig. 2(f) with Fig. 5(a)). We expect this
difference could reflect certain market information, however, the
underlying reason is not clear to us thus far.

\begin{ack}
This work is supported by the National Natural Science Foundation of
China under Grant Nos. 70571075 and 10635040, as well as the 973
Project 2006CB705500.
\end{ack}



\end{document}